# Current-induced synchronized switching of magnetization

Soo-Man Seo[1] and Kyung-Jin Lee[1,2,3,a]

[1]*Department of Materials Science and Engineering, Korea University, Seoul 136-701, Korea*

[2]*Center for Nanoscale Science and Technology, National Institute of Standards and Technology, Gaithersburg, Maryland 20899-8412, USA*

[3]*Maryland Nanocenter, University of Maryland, College Park, MD 20742, USA*

**We investigate current-induced magnetization switching for a multilayer structure that allows a reduced switching current while maintaining high thermal stability of the magnetization. The structure consists of a perpendicular polarizer, a perpendicular free-layer, and an additional free-layer having in-plane magnetization. When the current runs perpendicular to the structure, the in-plane free-layer undergoes a precession and supplies an internal rf field to the perpendicular free-layer, resulting in a reduced switching current for one current polarity. For the other polarity, the in-plane free-layer almost saturates perpendicular to the plane and acts as another perpendicular polarizer, which also reduces the switching current.**

[a] Electronic mail: kj_lee@korea.ac.kr.

Traditional charge-based memory technologies are approaching miniaturization limits as it becomes increasingly difficult to reliably retain sufficient electrons in the shrinking cells[1]. Because of this difficulty, non-volatile memories utilizing resistance change rather than charge storage have been investigated as potential alternatives to traditional charge-based memories. Such non-volatile memories include magnetic random access memory (MRAM)[2,3], phase-change memory[4], and resistive switching memory[5-7]. Among them, MRAM using current-induced magnetization switching as a write scheme has attracted considerable attention as a promising candidate for replacing dynamic random access memory beyond 30 nm lateral features because of its excellence in scalability, endurance, and speed.

Current-induced magnetization switching can be realized by the spin-transfer torque (STT) due to the coupling between local magnetic moments and spin-polarized currents[8,9]. Numerous studies on this subject have addressed its fundamental physics[10-16]. From the physics point of view, the basic mechanism of current-induced magnetization switching in multilayer is now well established[17,18] although intensive research is still ongoing for the complete understanding of STT physics[19-24]. From the application point of view, there are several critical issues to be resolved for realizing ultrahigh density STT-MRAM. One of these issues is how to keep decreasing the switching current as the magnetic cell gets smaller. The necessity to reduce the switching current is related to the size of transistor attached to the magnetic tunnel junction (MTJ) cell. If one wants to reduce the MTJ cell size (i.e., to increase memory density), one has to reduce the transistor size also. Since the current amplitude supplied by a transistor is limited by its lateral size, the switching current should be reduced accordingly. This issue is related to thermal stability of the magnetization which determines the retention time of stored information in a bit cell. The thermal stability is usually denoted by the thermal stability factor $\Delta$ ($= K_{eff}V/k_BT$) where $K_{eff}$ is the effective magnetic anisotropy, $V$ is the volume of magnetic cell, $k_B$ is the Boltzmann constant, and $T$ is the temperature. For 10-year data retention, $\Delta$ should exceed 50 at room temperature. In a perpendicular MTJ, which is superior to its in-plane counterpart for a higher density, both switching current and thermal stability factor are proportional to $K_{eff}V$ [25, 26]. To ensure an acceptable value of $\Delta$ for a smaller cell (= smaller $V$), $K_{eff}$ should increase accordingly, which in turn results in an increased switching current. Therefore, much effort has been expended on finding structures that provide a small switching current density while maintaining a high thermal stability factor. Another critical issue is that the switching current increases considerably as the current pulse gets shorter. This is because the switching dynamics changes from thermally activated switching to precessional switching as the pulse width becomes shorter[27]. Furthermore, the switching current distribution also increases at short pulses[28]. Note that in device applications, the driving current to switch the magnetization is set to a value corresponding to the switching probability of 100 %. Therefore, not only the average switching current but also the switching current distribution should be sufficiently small to maintain the driving current below an acceptable value.

In this Letter, we investigate current-induced magnetization switching for a multilayer structure that provides a smaller switching current while maintaining a high thermal stability factor, compared to the conventional standard structure. It also provides a narrower switching current distribution for short pulses. Figure 1 shows schematics of the standard perpendicular structure and the structure that we propose. The standard structure consists of a perpendicularly magnetized reference layer (p-RL; polarizer) and a perpendicular free layer (p-FL; switchable layer), separated by a nonmagnetic spacer (usually insulator). The proposed structure has an additional in-plane free layer (i-FL) on top of another spacer and p-FL. We name this structure "**C**urrent-**I**nduced **S**ynchronized **S**witching (CISS) structure" since the magnetization switching in this structure is assisted by synchronized magnetization dynamics due to an internally generated rf magnetic field as we will explain below. We note that CISS structure has some similarities to the previously studied STT oscillator consisting of an in-plane polarizer and a perpendicular free layer[29,30].

When the current runs perpendicular to the CISS structure, the magnetization of i-FL undergoes a precession motion due to the STT caused by p-FL. This precession motion in the structure consisting of a perpendicular layer, a spacer, and an in-plane layer was theoretically investigated by one of us[31] and proven experimentally[32]. Furthermore, for so-called orthogonal STT-MRAM[33,34], it has been experimentally demonstrated that perpendicular polarizer helps in-plane magnetization switch faster, governed by the same mechanism as for precession motion. The precession of i-FL supplies an rf magnetic field to p-FL. This internally driven rf field acts as an energy source, and thus reduces the switching current and its distribution. It is worthwhile comparing the CISS to previously reported works utilizing resonant effects. Cui *et al.*[35] and Florez *et al.*[36] experimentally demonstrated that an externally driven rf current in addition to a dc current reduces the switching current and its distribution. Carpentieri *et al.*[37] reported micromagnetic simulation results that an externally driven rf field in addition to a dc current also reduces the switching current at a given current pulse width. However, in these previous works, the rf current or field was *externally* driven so that an additional power consumption is inevitable. In contrast, in the CISS structure, the rf field is generated *internally* and thus an additional external source is not required, which makes the CISS attractive from the viewpoint of power consumption.

In this work, we investigate current-induced magnetization dynamics based on the coupled Landau-Lifshitz-Gilbert (LLG) equations including all in-plane STT terms as

$$\frac{d\mathbf{m}_p}{dt} = -\gamma\left(\mathbf{m}_p \times \mathbf{H}_p^{\text{eff}}\right) + \alpha\mathbf{m}_p \times \frac{d\mathbf{m}_p}{dt} - \frac{\gamma\hbar j_e \eta_0}{2eM_{S,p}d_p}\mathbf{m}_p \times \left(\mathbf{m}_p \times \hat{z}\right)$$

$$+ \frac{\gamma\hbar j_e \eta_{i\to p}}{2eM_{S,p}d_p}\mathbf{m}_p \times \left(\mathbf{m}_p \times \mathbf{m}_i\right) \quad (1)$$

$$\frac{d\mathbf{m}_i}{dt} = -\gamma\left(\mathbf{m}_i \times \mathbf{H}_i^{\text{eff}}\right) + \alpha\mathbf{m}_i \times \frac{d\mathbf{m}_i}{dt} - \frac{\gamma\hbar j_e \eta_{p\to i}}{2eM_{S,i}d_i}\mathbf{m}_i \times \left(\mathbf{m}_i \times \mathbf{m}_p\right)$$

where $\mathbf{m}$ is the unit vector along the magnetization, $\gamma$ is the gyromagnetic ratio (= $1.76\times10^{11}$ $T^{-1}s^{-1}$), $\mathbf{H}^{\text{eff}}$ is the effective magnetic field including the anisotropy, the dipolar, and the thermal fluctuation fields, $\alpha$ is the damping constant, $j_e$ is the current density, $M_S$ is the saturation magnetization, and $d$ is the layer thickness. Subscripts "$p$" and "$i$" correspond to p-FL and i-FL, respectively, and $\eta_0$, $\eta_{p\to i}$, and $\eta_{i\to p}$ are the efficiency factors of STT from p-RL to p-FL, from p-FL to i-FL, and from i-FL to p-FL, respectively. To get an insight into the CISS, in Eq. (1), we neglect perpendicular STT terms that can be large in MgO-based MTJs[19-24].

Here we present 3-dimensional micromagnetic simulation with the following parameters; $\eta_0 = \eta_{p\to i} = \eta_{i\to p} = 0.5$, $M_{S,p}$ = 1100 kA/m, $\alpha$ = 0.01, the magnetic anisotropy energy density $K_{u,p} = 7\times10^5$ J/m$^3$, $K_{u,i}$ = 0, the exchange constant $A_{ex,p} = 2\times10^{-11}$ J/m, $A_{ex,i} = 1.3\times10^{-11}$ J/m, $d_p$ = 3 nm, $d_i$ = 1 nm, the thickness of spacer between p-FL and i-FL of 1 nm, a circular cross-section of the nanopillar with a diameter of 20 nm, and no dipolar field from p-RL. The unit cell of $2\times2\times d_p$ ($d_i$) nm$^3$ was used and the current-induced Oersted field was considered.

Figure 2(a) and (b) show the time evolution of the magnetizations in p-FL and i-FL during the switching. For the parallel-to-antiparallel (P-to-AP) switching (Fig. 2(a)), the in-plane ($x$ or $y$) components of the magnetizations in p-FL and i-FL are exactly out of phase by $\pi$, indicating that the dynamics of the two magnetizations are synchronized by the coupling through STTs and dipolar fields. The out-of-plane components of the magnetizations ($m_z$) in both layers switch while maintaining synchronized dynamics. It is interesting to observe that the precession frequency is about 8 GHz (for instance, see the top panel of Fig. 2(a); 16 oscillation peaks during the initial 2 ns), which is much lower than the Lamor frequency of p-FL (≈ 22 GHz). We note that for purely field-driven magnetization switching, a reduced coercive field assisted by an external rf field with the frequency below the Lamor frequency was reported[38,39]. Even after the switching of $m_z$, the synchronized dynamics is still maintained and results in a steady precession motion. However, we find that when the current is turned off after the perpendicular switching of both layers, this steady precession motion does not cause a switching-back phenomenon.

On the other hand, the magnetization dynamics for the AP-to-P switching is quite different (Fig. 2(b)). In the initial time stage, the magnetization of i-FL almost saturates along the perpendicular (= thickness) direction. This is because both the STT and the stray field from p-FL to i-FL encourage the out-of-plane saturation of i-FL. The magnetic configuration then becomes similar to that of the double spin-filter proposed by Berger[40]; both p-RL and i-FL have strong out-of-plane components with opposite direction. As a result, the STT efficiency experienced by p-FL is increased in comparison to the standard structure[3,38].

Reduction in the switching current for the AP-to-P switching in the CISS structure is obvious because of the double spin-filter effect. Here we try to derive an analytical threshold current for the P-to-AP switching in the CISS structure. Since the two magnetizations in p-FL and i-FL are nonlinearly coupled through not only stray fields but also STTs, it would be rather difficult to get a simple analytical solution. To simplify the problem, we assume; i) angular frequencies of p-FL and i-FL are the same as $\omega$ because of the synchronization, ii) the magnetization of i-FL stays in the plane, and iii) STT from i-FL to p-FL is zero ($\eta_{i \to p} = 0$). With these approximations, the problem is now reduced to the case where both a torque due to the rotating in-plane magnetic field with the frequency of $\omega/2\pi$ and a STT from p-RL are simultaneously exerted on the magnetization of p-FL. With $\mathbf{m}_p = (\cos\omega t \sin\theta, \sin\omega t \sin\theta, \cos\theta)$ where $\theta$ is the polar angle of $\mathbf{m}_p$, one can get the LLG equations in the spherical coordinate system;

$$\omega \sin\omega t \sin\theta - \cos\omega t \cos\theta \frac{d\theta}{dt} = \gamma\left(H_k^{eff} \sin\omega t \sin\theta \cos\theta - H_{rf} \sin\omega t \cos\theta\right)$$
$$+ \alpha\left(\sin\omega t \frac{d\theta}{dt} + \omega \cos\omega t \cos\theta \sin\theta\right) \quad (2)$$
$$- \gamma a_J \cos\omega t \cos\theta \sin\theta$$

$$\frac{d\theta}{dt} = -(\alpha\omega + \gamma a_J)\sin\theta \quad (3)$$

where $H_k^{eff}$ is the effective anisotropy field of p-FL including both crystalline anisotropy and demagnetization fields, $H_{rf}$ is the magnitude of rf field generated by i-FL, and $a_J = \hbar j_e \eta_0 / 2eM_{S,p}d_p$. For a synchronized steady-state precession mode ($d\theta/dt = 0$), one finds the solution

$$\omega \equiv \gamma|a_J|/\alpha = \gamma\left(H_k^{eff}\cos\theta - H_{rf}\cot\theta\right) \quad (4)$$

The threshold $a_J$ (= $a_c$) can be obtained from the stability condition of Eq. (4). In other words, the solution becomes unstable above a threshold current so that the steady-state precession ($d\theta/dt = 0$) is not maintained anymore and the magnetization of p-FL eventually switches. Since Eq. (4) implies a linear increase in $\omega$ with the current, $a_c$ is obtained from a critical $\theta$ (= $\theta_c$) that maximizes the right-hand-side of Eq. (4). Then, one finds $\theta_c$ and $a_c$ as

$$\theta_c = \cos^{-1}\left(\sqrt{1 - \left(H_{rf}/H_k^{eff}\right)^{2/3}}\right), \quad (5)$$

$$a_c = \alpha\left(H_k^{eff}\cos\theta_c - H_{rf}\cot\theta_c\right) \quad (6)$$

Note that Eq. (6) can be also derived from the switching condition ($d\theta/dt > 0$) in Eqs. (3)

and (4). It is known that the threshold $a_c$ for the standard structure is $\alpha H_k^{eff}$ [25]. According to Eq. (6), it is obvious that the CISS with $H_{rf} \neq 0$ has a lower $a_c$ than the standard structure. For instance, with the parameters used for Fig. 2(a), we find $\alpha H_k^{eff}$ = 4.2 mT and $a_c$ = 3.3 mT, which gives about 20 % reduction in the switching current.

Eq. (6) is derived based on several approximations so that it may not be used to quantitatively estimate the switching current of the CISS structure. Figure 2(c) shows power spectra obtained from Fourier transformation of $m_x$ for the i-FL and p-FL with the time window of 10 ns and $T$ = 0 K, respectively. The dotted white lines indicate the switching thresholds. As predicted analytically, the frequency of i-FL linearly increases with the current when the current is small. For positively large current (> 10μA), however, the precession frequency does not follow the linear dependence because of the synchronization.

For a quantitative estimation, therefore, we numerically solve Eq. (1) with the saturation magnetization of i-FL as a variable. Average switching current and thermal stability factor as a function of the saturation magnetization of i-FL are summarized in Fig. 3(a) and 3(b), respectively. The average switching current is determined as the current for the switching probability of 50 % at $T$ = 300 K and pulse width = 10 ns. The thermal stability factor is obtained from the total energy calculation. Black dashed lines in Fig. 3(a) and 3(b) represent the average switching current and the thermal stability factor of the standard structure, respectively. For both P-to-AP and AP-to-P switchings, the CISS structure exhibits a reduced switching current by about 40 % compared to the standard structure. As explained above, the switching current for the P-to-AP switching is reduced by the synchronized dynamics whereas that for the AP-to-P switching is reduced by the double spin-filter effect. On the other hand, the thermal stability factor is reduced by about 10 % at most in the tested ranges of the saturation magnetization of i-FL. As a result, the ratio of the thermal stability factor to the switching current is improved when adopting the CISS structure.

Another advantage of the CISS structure is a narrow switching current distribution. Figure 4(a) shows the switching probability ($P_{SW}$) versus the applied current ($I_{app}$). $P_{SW}$ is obtained from the number of successful switching events out of 100 trials at each current. The switching probability curves are fitted by the cumulative distribution function $P_{SW} = \{1 + \mathrm{erf}[(I_{app} - I_{SW})/(\sigma\sqrt{2})]\}/2$ where $I_{SW}$ is the average switching current and $\sigma$ is the standard deviation of the switching current. Figure 4(b) shows $d(P_{SW})/(dI_{app})$ for three cases; the standard structure, the P-to-AP and AP-to-P switchings of the CISS structure. It is evident that the CISS structure yields narrower switching current distributions than the standard structure. Furthermore, the P-to-AP switching (CISS structure) exhibits the narrowest switching current distribution.

Figure 4(c) and 4(d) show the average switching current ($I_{SW}$) and the standard deviation ($\sigma$) as a function of the pulse width ($\tau$), respectively. It is known that both $I_{SW}$ and $\sigma$ rapidly increase with decreasing $\tau$ because the switching dynamics changes from thermally activated switching to precessional switching. Interestingly, the P-to-AP switching of the CISS structure exhibits a smaller increase in $I_{SW}$ and $\sigma$ than other two cases. We attribute this smaller increase to synchronized precessional switching. It is because the P-to-AP switching of the CISS structure is always achieved through a kind of precessional switching regardless of the pulse width $\tau$. Therefore, there is no transition from thermally activated switching to precession switching in this case, which makes both $I_{SW}$ and $\sigma$ less sensitive to $\tau$. This explanation is also supported by the variation of the ratio of $\sigma$ to $I_{SW}$ as a function of $\tau$ (the inset of Fig. 4(d)). The ratio for the P-to-AP switching of the CISS structure does not vary much with $\tau$, whereas the ratio for the AP-to-P switching of the CISS structure or for the switching of the standard structure substantially increases with decreasing $\tau$. This lack of abrupt increase in the switching current at short pulses for the P-to-AP switching would be beneficial for the device applications since the driving current is usually set to the switching current for the P-to-AP switching that is larger than that for the AP-to-P switching in MTJs.

To conclude, we investigate current-induced magnetization switching in a multilayer structure, the CISS structure, consisting of an additional i-FL/spacer (or insulator) on top of the standard all perpendicular MTJ. The CISS structure provides several advantages over the standard structure; i) an improved ratio of the thermal stability factor to the switching current, ii) a narrower switching current distribution, iii) no abrupt increase in the switching current for P-to-AP switching at short pulses. We also remark that the CISS structure would be more advantageous than the double spin-filter structure consisting of p-RL1|spacer1|p-FL|spacer2|p-RL2 where the perpendicular magnetizations of p-RL1 and p-RL2 are opposite with each other. At the remnant state, the resistance change between the P and AP states in the double spin-filter structure is smaller than that of the standard structure because of the opposite orientation of two p-RLs. However, in the CISS structure, the resistance change could be similar to that of the standard structure since the magnetization of i-FL stays mostly in the plane at the remnant state. Thus, the CISS structure will be beneficial for ultrahigh density magnetic memories and logic devices.

This work was supported by the NRF (2010-0023798) the MKE/KEIT (2009-F-004-01). K. J. Lee acknowledges support under the Cooperative Research Agreement between the University of Maryland and the National Institute of Standards and Technology Center for Nanoscale Science and Technology, Award 70NANB10H193, through the University of Maryland.

**REFERENCE**

[1] G. I. Meijer, Science **319**, 1625 (2008).

[2] S. Ikeda, J. Hayakawa, Y. M. Lee, F. Matsukura, Y. Ohno, T. Hanyu, and H. Ohno, IEEE Trans. Electron Devices **54**, 991 (2007).

[3] Z. Diao, A. Panchula, Y. Ding, M. Pakala, S. Wang, Z. Li, D. Apalkov, H. Nagai, A. Driskill-Smith, L.-C. Wang, E. Chen, and Y. Huai, Appl. Phys. Lett. **90**, 132508 (2007).

[4] S. Lai, IEDM Tech. Dig. 255 (2003).

[5] A. Beck, J. G. Bednorz, C. Gerber, C. Rossel, D. Widmer, Appl. Phys. Lett. **77**, 139 (2000).

[6] S.-J. Choi, G.-B. Kim, K. Lee, K.-H. Kim, W.-Y. Yang, S. Cho, H.-J. Bae, D.-S. Seo, S.-I. Kim, and K.-J. Lee, Appl. Phys. A **102**, 1019 (2011).

[7] S.-J. Choi, G.-S. Park, K.-H. Kim, S. Cho, W.-Y. Yang, X.-S. Li, J.-H,. Moon, K.-J. Lee, K. Kim, Adv. Mater. **23**, 3272 (2011).

[8] J. C. Slonczewski, J. Magn. Mag. Mater. **159**, L1 (1996).

[9] L. Berger, Phys. Rev. B **54**, 9353 (1996).

[10] J. A. Katine, F. J. Albert, R. A. Buhrman, E. B. Myers, and D. C. Ralph, Phys. Rev. Lett. **84**, 3149 (2000).

[11] J. Grollier, V. Cros, A. Hamzic, J. M. George, H. Jaffres, A. Fert, G. Faini, J. B. Youssef, and H. Legall, Appl. Phys. Lett. **78**, 3663 (2001).

[12] M. D. Stiles and A. Zangwill, Phys. Rev. B **66**, 014407 (2002).

[13] Y. Tserkovnyak, A. Brataas, and G. E. W. Bauer, Phys. Rev. Lett. **88**, 117601 (2002).

[14] S. Zhang, P. M. Levy, and A. Fert, Phys. Rev. Lett. **88**, 236601 (2002).

[15] S. Urazhdin, N. O. Birge, W. P. Pratt, Jr., and J. Bass, Phys. Rev. Lett. **91**, 146803 (2003).

[16] K. J. Lee, Y. Liu, A. Deac, M. Li, J. W. Chang, S. Liao, K. Ju, O. Redon, J. P. Nozières, and B. Dieny, J. Appl. Phys. **95**, 7423 (2004).

[17] D. C. Ralph and M. D. Stiles, J. Magn. Magn. Mater. **320**, 1190 (2008).

[18] J. Z. Sun and D. C. Ralph, J. Magn. Magn. Mater. **320**, 1227 (2008).

[19] H. Kubota, A. Fukushima, K. Yakushiji, T. Nagahama, S. Yuasa, K. Ando, H. Maehara, Y. Nagamine, K. Tsunekawa, D. D. Djayaprawira, N. Watanabe, and Y. Suzuki, Nature Phys. **4**, 37 (2008).

[20] J. C. Sankey, Y.-T. Cui, J. Z. Sun, J. C. Slonczewski, R. A. Buhrman, and D. C. Ralph, Nature Phys. **4**, 67 (2008).

[21] S.-C. Oh, S.-Y. Park, A. Manchon, M. Chshiev, J.-H. Han, H.-W. Lee, J.-E. Lee, K.-T. Nam, Y. Jo, Y.-C. Kong, B. Dieny, and K.-J. Lee, Nature Phys. **5**, 898 (2009).

[22] O. G. Heinonen, S. W. Stokes, and J. Y. Yi, Phys. Rev. Lett. **105**, 066602 (2010).

[23] C. Wang, Y.-T. Cui, J. A. Katine, R. A. Buhrman, and D. C. Ralph, Nature Phys. **7**, 496 (2011).

[24] S.-Y. Park, Y. Jo, and K.-J. Lee, Phys. Rev. B **84**, 214417 (2011).

[25] S. Mangin, D. Ravelosona, J. A. Katine, M. J. Carey, B. D. Terris, and E. E. Fullerton, Nature Mater. **5**, 210 (2006).

[26] S. Ikeda, K. Miura, H. Yamamoto, K. Mizunuma, H. D. Gan, M. Endo, S. Kanai, J. Hayakawa, F. Matsukura, and H. Ohno, Nature Mater. **9**, 721 (2010).

[27] D. Bedau, H. Liu, J. Z. Sun, J. A. Katine, E. E. Fullerton, S. Mangin, and A. D. Kent, Appl. Phys. Lett. **97**, 262502 (2010).

[28] W.-Y. Kim and K.-J. Lee, J. Kor. Magn. Soc. **21**, 48 (2011).

[29] Y. Zhou, C. L. Zha, S. Bonetti, J. Persson, and J. Akerman, Appl. Phys. Lett. **92**, 262508 (2008).

[30] W. H. Rippard, A. M. Deac, M. R. Pufall, J. M. Shaw, M. W. Keller, S. E. Russek, G. E. W. Bauer, and C. Serpico, Phys. Rev. B **81**, 014426 (2010).

[31] K. J. Lee, O. Redon, and B. Dieny, Appl. Phys. Lett. **86**, 022505 (2005).

[32] D. Houssameddine, U. Ebels, B. Delaët, B. Rodmacq, I. Firastrau, F. Ponthenier, M. Brunet, C. Thirion, J.-P. Michel, L. Prejbeanu-Buda, M.-C. Cyrille, O. Redon, and B. Dieny, Nature Mater. **6**, 447 (2007).

[33] A. D. Kent, B. Özyilmaz, and E. del Barco, Appl. Phys. Lett. **84**, 3897 (2004).

[34] O. J. Lee, D. C. Ralph, and R. A. Buhrman, Appl. Phys. Lett. **99**, 102507 (2011).

[35] Y.-T. Cui, J. C. Sankey, C. Wang, K. V. Thadani, Z.-P. Li, R. A. Buhrman, and D. C. Ralph,

Phys. Rev. B **77**, 214440 (2008).

[36] S. H. Florez, J. A. Katin, M. Carey, L. Folks, O. Ozatay, and B. D. Terris, Phys. Rev. B **78**, 184403 (2008).

[37] M. Carpentieri, G. Finocchio, B. Azzerboni, and L. Torres, Phys. Rev. B **82**, 094434 (2010).

[38] Z. Z. Sun and X. R. Wang, Phys. Rev. B **74**, 132401 (2006).

[39] J.-G. Zhu, X. Zhu, and Y. Tang, IEEE Trans. Magn. **44**, 125 (2008).

[40] L. Berger, J. Appl. Phys. **93**, 7693 (2003).

**FIGURE CAPTION**

FIG. 1. (Color online) Schematics of (a) standard structure and (b) CISS structure of perpendicular STT-MRAM.

FIG. 2. (Color online) Magnetization switching in the CISS structure at $T = 0$ K: Normalized magnetization vector (**m**) as a function of the time for (a) P-to-AP switching with $I_{app} = +30$ µA and (b) AP-to-P switching with $I_{app} = -30$ µA. The saturation magnetization of free layer with in-plane anisotropy (i-FL) is 800 kA/m. (c) Frequency power spectra of in-plane free layer and perpendicular free layer, obtained from Fourier transformation of $m_x$ with the time window of 10 ns and $T = 0$K. The dotted white lines indicate the switching thresholds.

FIG. 3. (Color online) (a) Switching current as a function of the saturation magnetization ($M_S$) of free layer with in-plane anisotropy (i-FL). All calculations have been done at $T = 300$ K. (b) Thermal stability factor ($= K_{eff}V/k_BT_{300}$) as a function of $M_S$ of i-FL.

FIG. 4. (Color online) Switching current distribution: (a) Switching probability ($P_{SW}$) as a function of applied current ($I_{app}$). (b) $dP_{SW}/dI_{app}$ as a function of $I_{app}$. (c) Switching current ($I_{SW}$) as a function of the current pulse width ($\tau$). (d) Switching current distribution ($= \sigma$) as a function of $\tau$. In (c) and (d), black open square symbols correspond to the standard structure whereas red open circle (blue open up triangle) symbols correspond to the P-to-AP (AP-to-P) switching of the CISS structure. The inset of (d) shows the ratio of $\sigma$ to $I_{SW}$ as a function of $\tau$.

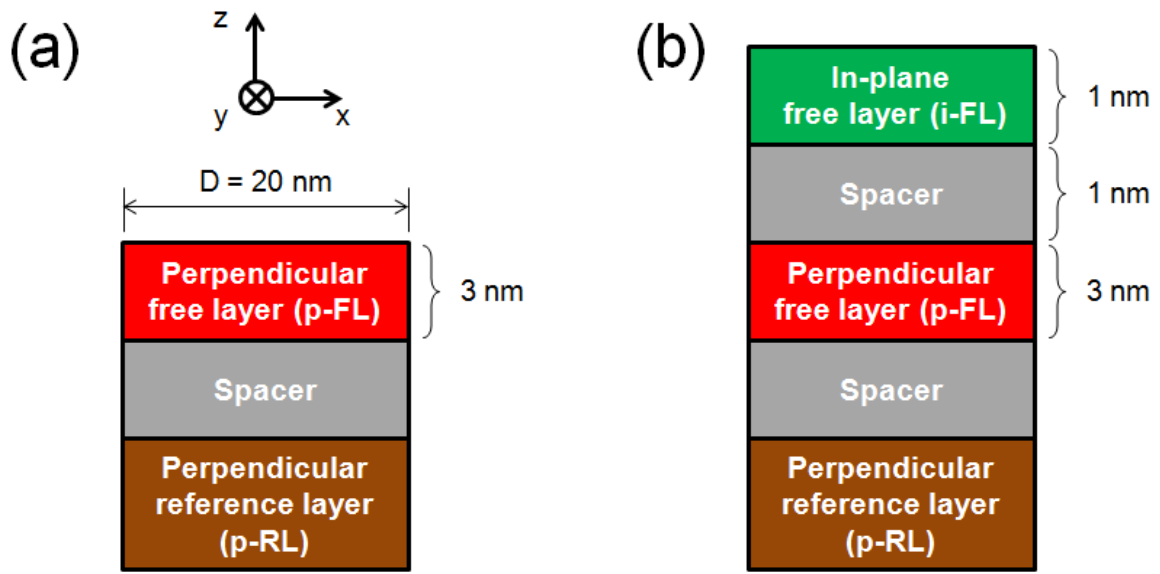

FIG. 1. Seo *et al*.

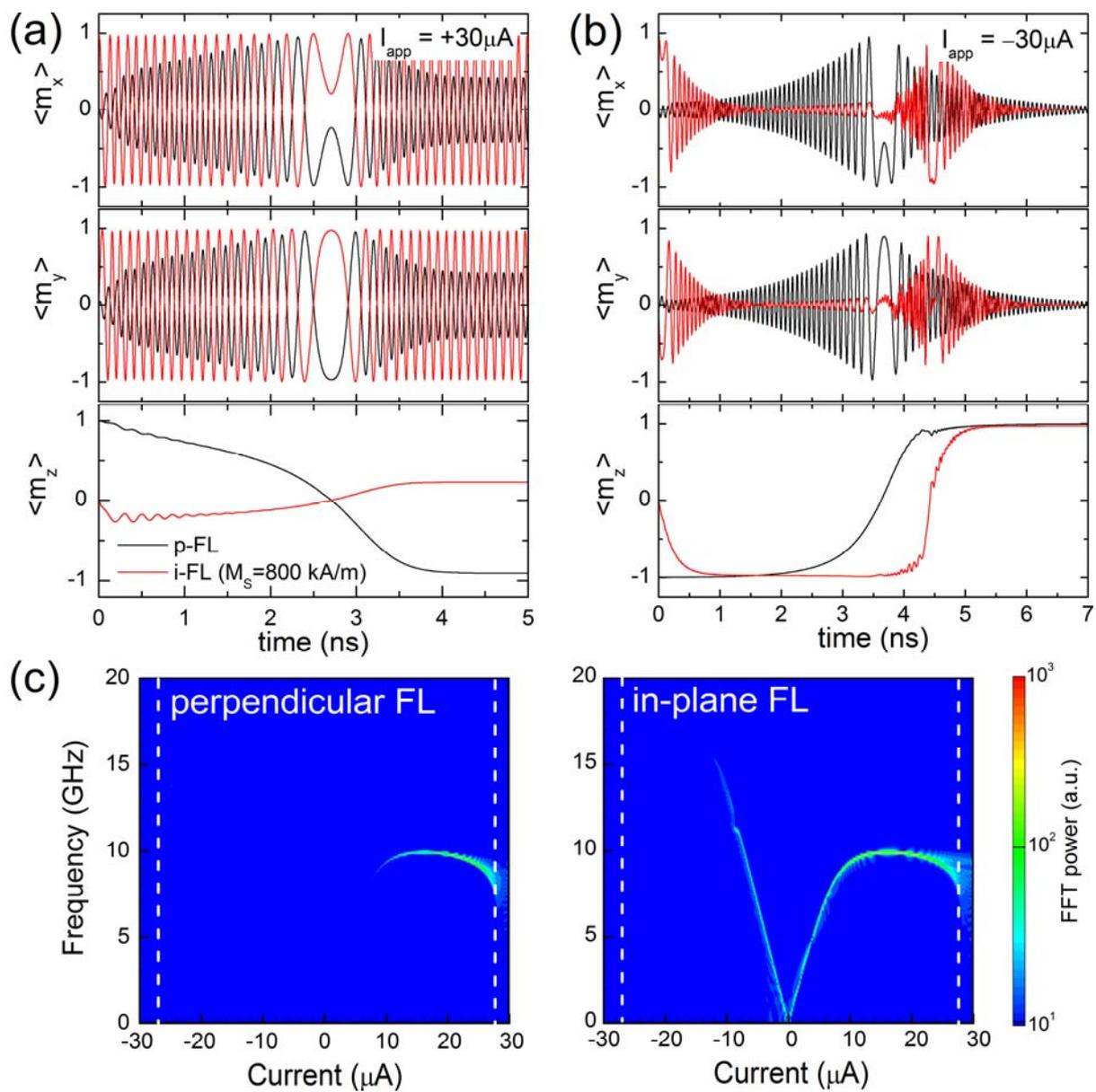

FIG. 2. Seo *et al*.

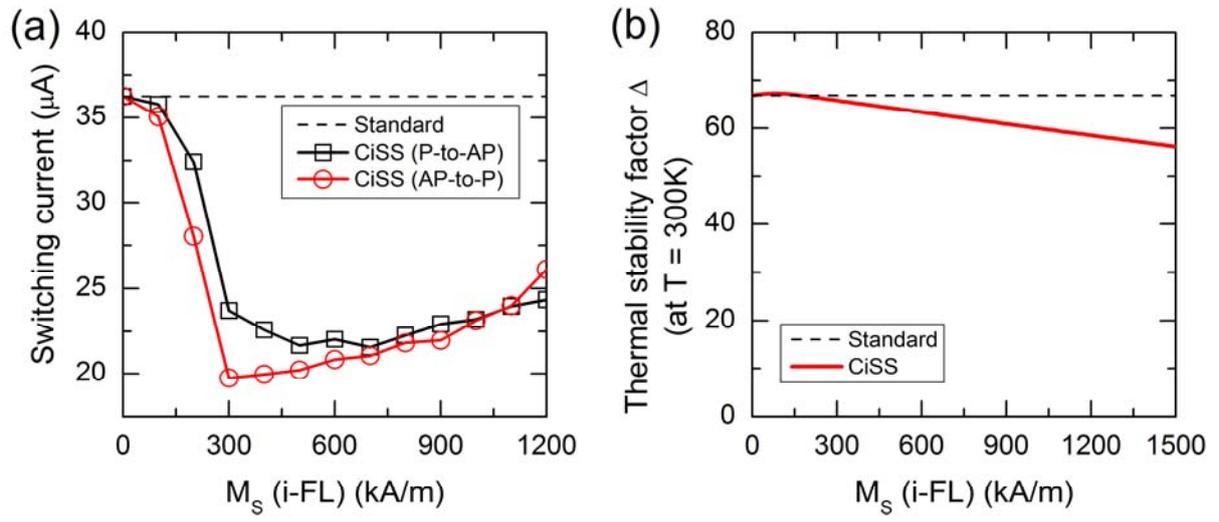

FIG. 3. Seo *et al*.

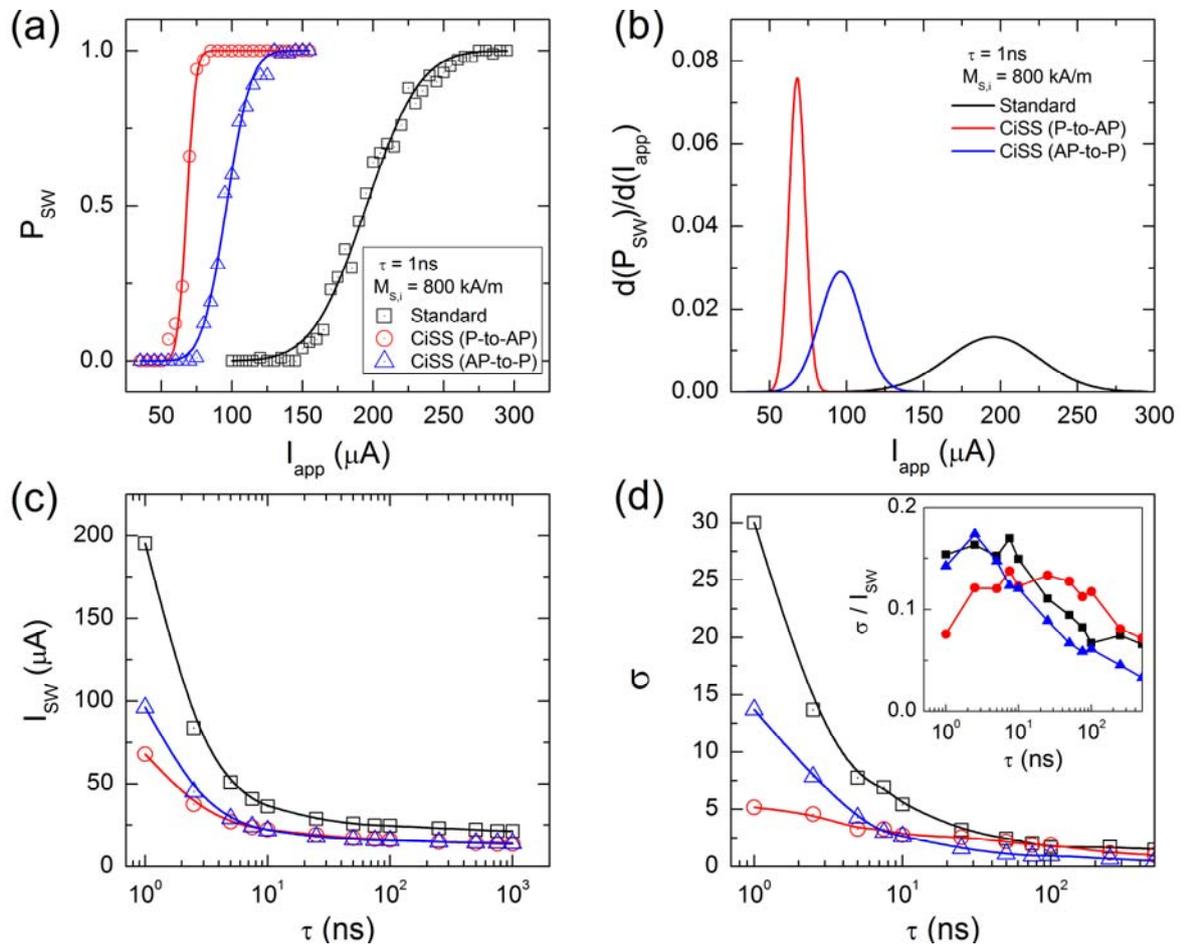

FIG. 4. Seo *et al*.